\documentclass[10pt]{iopart}
\pdfoutput=1

\expandafter\let\csname equation*\endcsname\relax

\expandafter\let\csname endequation*\endcsname\relax

\usepackage{amsmath}

\usepackage[utf8]{inputenc}
\usepackage{multirow}

\usepackage[dvips]{graphics,graphicx}
\usepackage{url}
\usepackage{subfigure}
\usepackage{etoolbox}
\usepackage{booktabs}
\usepackage[table,xcdraw]{xcolor}
\usepackage{ulem}
\usepackage{overpic}
\usepackage{hyperref}

\usepackage{color}

\newcommand{\dd}[2]{\frac{\partial #1}{\partial #2}}
\newcommand{\tc}{\tau_c}
\newcommand{\tcorr}{\tau_\mathrm{corr}}
\newcommand{\DRR}{D_\mathrm{RR}}
\newcommand{\Dc}{D_c}
\newcommand{\Dnum}{D_\mathrm{NUM}}
\newcommand{\rhopol}{\rho}
\newcommand{\rhoedge}{\rho_\mathrm{pass}}
\newcommand{\jacobian}{J}
\newcommand{\driftK}{K}

\newcommand{\diffusionD}{D}
\newcommand{\pitch}{\xi}
\newcommand{\energy}{E}
\newcommand{\wiener}{\mathcal{W}}
\newcommand{\Ito}{It\^o}
\newcommand{\Peclet}{P\'eclet}
\newcommand{\Poincare}{Poincar\'e}
\newcommand{\separation}{\mathbf{X}}
\newcommand{\noisedrift}{K_D}

\begin{document}

\title[An advection-diffusion model for cross-field runaway electron transport in perturbed magnetic fields]{An advection-diffusion model for cross-field runaway electron transport in perturbed magnetic fields}

\author{Konsta Särkimäki\textsuperscript{1},
  Eero Hirvijoki\textsuperscript{2},
  Joan Decker\textsuperscript{3},
  Jari Varje\textsuperscript{1},
  Taina Kurki-Suonio\textsuperscript{1,4}}

\address{\textsuperscript{1}Aalto University, Espoo, Finland}
\address{\textsuperscript{2}PPPL, Princeton, USA}
\address{\textsuperscript{3}EPFL, Lausanne, Switzerland}
\address{\textsuperscript{4}Chalmers Technical University, Gothenburg, Sweden}

\ead{konsta.sarkimaki@aalto.fi}
\vspace{10pt}
\begin{indented}
\item[]\today
\end{indented}

\begin{abstract}\\
Disruption-generated runaway electrons (RE) present an outstanding issue for ITER. The predictive computational studies of RE generation rely on orbit-averaged computations and, as such, they lack the effects from the magnetic field stochasticity. Since stochasiticity is naturally present in post-disruption plasma, and externally induced stochastization offers a prominent mechanism to mitigate RE avalanche, we present an advection-diffusion model that can be used to couple an orbit-following code to an orbit-averaged tool in order to capture the cross-field transport and to overcome the latter's limitation. The transport coefficients are evaluated via a Monte Carlo method. We show that the diffusion coefficient differs significantly from the well-known Rechester-Rosenbluth result. We also demonstrate the importance of including the advection: it has a two-fold role both in modelling transport barriers created by magnetic islands and in amplifying losses in regions where the islands are not present.

\end{abstract}

%
\vspace{2pc}

%

%
%
\ioptwocol

\section{Introduction}
\label{sec:Introduction}

Based on analytical work and reduced numerical models, existence of a large runaway electron (RE) population is anticipated in ITER during plasma terminating events~\cite{rosenbluth1997theory,eriksson2004current}. According to the predictions, majority of the initial plasma current is expected to be converted into runaway current via an avalanche process, possibly endangering the entire experiment~\cite{0029-5515-49-6-065012,hender2007mhd}. One proposed remedy is to introduce resonant magnetic perturbations (RMP) that  create a layer of stochastic magnetic field lines at the plasma edge~\cite{helander2000suppression}. Since runaway electrons are both strongly passing and practically collisionless, the layer would enhance the cross-field transport, which in turn could lead to losses that avert the runaway avalanche. Assessing the feasibility of this scheme would require 5D time-dependent drift-kinetic calculations involving realistic geometry and effects such as island formation. (Even without RMPs, studies of post-disruption runaways should involve these features as the field becomes stochastic during the thermal quench.) Because the 5D computations are so demanding, we study here an alternative method where continuum-based orbit-averaged tools are coupled to particle-based orbit-following codes in order to efficiently capture both the generation of runaways and their redistribution in a realistic magnetic field. To this end, we describe the implementation of a 1D transport model for runaways, and demonstrate that it captures the radial transport that is caused by the field perturbations. Future work is to couple the model into an orbit-averaged tool.

\begin{figure}[!t]
\centering
\begin{overpic}[width=0.45\textwidth]{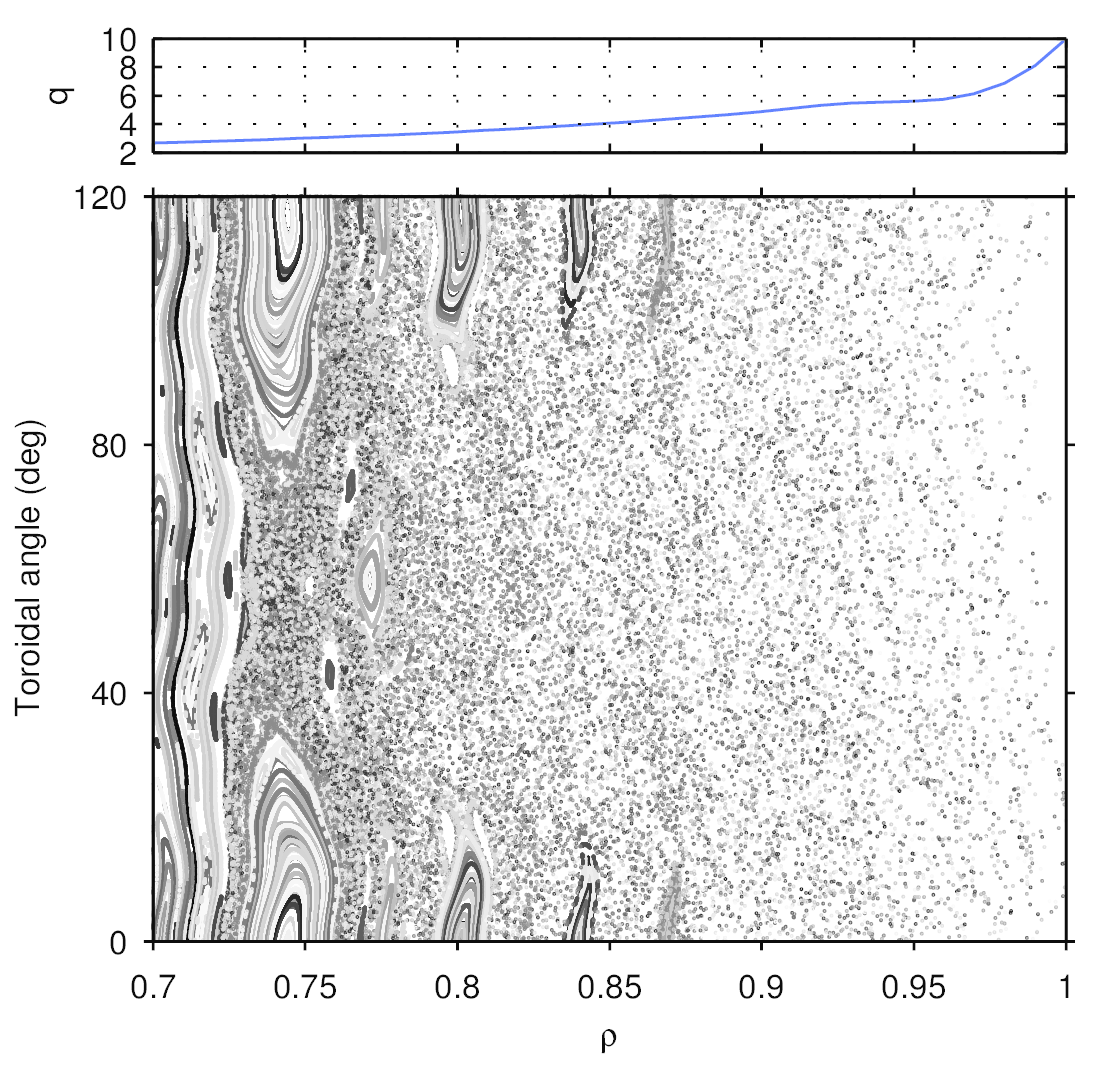}
\end{overpic}
\caption{The $q$-profile (top) and \Poincare{} plot showing the magnetic field structure at outer midplane. The field consists of 2D equilibrium, corresponding to ITER $I_p = 9$ MA non-inductive scenario, and toroidal ripple and is perturbed with RMP coils. The coils are in a current configuration where mode is $n=3$, and amplitude is 95 kAt.  Due to $n=3$ symmetry, only one sector is shown. The field lacks plasma response as it was solely constructed to study RE transport in this work.}
\label{fig:poincare9MA}
\end{figure}

We resolve the transport coefficients involved in the model via Monte Carlo evaluation of RE transport with an orbit-following code, because the particle transport in a stochastic field is not yet well-understood. Even though the particle motion itself is simply given by the Lorentz force law, the chaotic nature of the field leads to a diffusive transport of the population as a whole~\cite{rosenbluth1966destruction}. The mean square displacement during one poloidal orbit is related to the magnetic field perturbation as
\begin{equation}
\label{eq:RR mean square displacement}
\left<(\Delta x)^2\right> \propto \sigma^2\tilde{b}^2,
\end{equation}
where $\sigma$ is the orbit length and $\tilde{b}^2\equiv \langle\left(\delta B / B \right)^2\rangle$ is the orbit-average of perturbation amplitude normalized to unperturbed field magnitude. For a particle with parallel velocity $v_\parallel$ the circulation time is $\tau_\mathrm{orb} = \sigma / v_\parallel$, which leads to the \emph{Rechester-Rosenbluth diffusion coefficient}~\cite{rechester1978electron}
\begin{equation}
\label{eq:RR diffusion coeff}
\frac{\left<(\Delta x)^2\right>}{2\tau} = C\frac{v_\parallel\sigma}{2}\tilde{b}^2 \equiv \DRR,
\end{equation}
where $C$ is a constant. However, relation~\eqref{eq:RR mean square displacement} is based on the assumption that no remnant islands are present, and particles have zero orbit widths. The perturbed magnetic fields in tokamaks tend to have complex structure as illustrated in Fig.~\ref{fig:poincare9MA}. Between the healthy flux surfaces ($\rhopol < 0.72 $) and the very stochastic edge region ($\rhopol > 0.88 $), which appears to have no structure left, lies a region where stochastic volume is dotted with islands of varying size. The islands are known to reduce the diffusion below levels predicted by Eq.~\eqref{eq:RR diffusion coeff}, creating transport barriers near rational surfaces~\cite{abdullaev2012new}. The diffusion coefficient is also reduced when REs have high energy ($\energy_\mathrm{kin} > $ 10 MeV) and the orbit width effects become non-negligible~\cite{myra1993quasilinear}. Therefore, we cannot rely on Eq.~\eqref{eq:RR diffusion coeff} to describe the transport correctly. 

Different models for the diffusion coefficient has been developed and compared with numerical values~\cite{abdullaev2012new,myra1993quasilinear,abdullaev2014magnetic,hauff2009runaway,martin1999effect} but, as was shown in Ref.~\cite{papp2015energetic} using orbit-following tools, the transport is not purely diffusive. Furthermore, islands have been argued to exert friction on particles~\cite{sugimoto1996friction}. In this work, we present an advection-diffusion model, and show that it is able to capture the RE transport, thus confirming the presence of an advection process. Section 2 presents the 1D-transport equation and discusses methods to solve it numerically. The Monte Carlo evaluation of the transport coefficients is detailed in Section 3 and the coefficients are then evaluated for the magnetic field presented in Fig.~\ref{fig:poincare9MA}. There it is also explicitly shown how the numerical diffusion coefficient differs from the expression~\eqref{eq:RR diffusion coeff} and how Eq.~\eqref{eq:RR diffusion coeff} can be modified to obtain a better agreement with the numerical values. Using the Monte Carlo evaluated advection and diffusion coefficients, we also show in Section 4 that the 1D-model is able to reproduce the RE transport obtained with an orbit-following simulation. Section 5 contains analysis of the relation between RE parameters and the transport coefficients. The paper is concluded in Section 6.

\section{Advection-diffusion model}
\label{sec:Drift-diffusion model}

Orbit-averaged codes designed for RE studies, such as LUKE~\cite{decker2004dke}, operate in a phase space $(x,\pitch,\energy)$, where $x$ is the radial coordinate, and $\pitch \equiv v_\parallel/v$ and $\energy$ are the particle pitch and energy measured when the particle crosses the outer midplane (OMP). Since the energy and pitch are evolved under the orbit-averaged formalism, the transport model should not change these coordinates. This means that electric field acceleration and collisions must not be accounted for during the poloidal orbit the particle is followed. If radial transport occurs during the poloidal orbit, the adiabatic conservation of the particle's magnetic moment will affect the pitch but, since the variation in the field magnitude on OMP is small for small radial displacements, the change in the pitch is negligible. The transport model is thus effectively one-dimensional, and parametrized by $\pitch$ and $\energy$. As a radial coordinate, $x$, we use $\rhopol = \sqrt{(\psi-\psi_\mathrm{axis})/(\psi_\mathrm{sep}-\psi_\mathrm{axis})}$, where $\psi_\mathrm{axis}$ and $\psi_\mathrm{sep}$ are the poloidal fluxes at the magnetic axis and separatrix, respectively. We consider a transport model that includes an advection term in addition to diffusion, so that the radial density, $f(\rhopol,\pitch,\energy,t)$, evolves according to the \emph{Fokker-Planck equation}
\begin{align}
\label{eq:Fokker-Planck}
\dd{f(\rhopol,\pitch,\energy,t)}{t} &= -\frac{1}{\jacobian}\dd{}{\rhopol}\left[\jacobian\driftK(\rhopol,\pitch,\energy) f(\rhopol,\pitch,\energy,t)\right]\nonumber\\
&+\frac{1}{\jacobian}\dd{^2}{\rhopol^2}\left[\jacobian\diffusionD(\rhopol,\pitch,\energy) f(\rhopol,\pitch,\energy,t)\right],
\end{align}
where $\driftK$ and $\diffusionD$ are advection (or drift) and diffusion coefficients defined as 
\begin{align}
\label{eq:definition of K}
\driftK &\equiv \lim_{\Delta t \rightarrow 0}\frac{\left\langle \Delta \rhopol \right\rangle}{\Delta t},\\
\label{eq:definition of D}
\diffusionD &\equiv \lim_{\Delta t \rightarrow 0}\frac{\left\langle (\Delta \rhopol)^2 \right\rangle}{\Delta t}.
\end{align}
Here brackets denote ensemble average and $\Delta \rhopol$ is radial displacement during time interval $\Delta t$. We may drop the $\energy$ and $\pitch$ labels since those stay constant in our simulations, and we can treat populations with different $\energy$ and $\pitch$ separately. The Jacobian in Eq.~\eqref{eq:Fokker-Planck} is $\jacobian = \partial R/\partial\rhopol$ where $R$ is the major radius at OMP.

Before proceeding further in developing the transport model, we briefly discuss the Fokker-Planck equation in general. The connection between the Fokker-Planck equation and the stochastic differential equation (SDE) known as the Langevin equation, is well known: The probability density of a random variable obeying the Langevin equation evolves according to the Fokker-Planck equation. In other words, if RE density evolves according to the Eq.~\eqref{eq:Fokker-Planck}, the radial position of each individual RE is governed by the Langevin equation. In \emph{\Ito{} convention}, the Langevin equation corresponding to the Eq.~\eqref{eq:Fokker-Planck} is
\begin{equation}
\label{eq:Langevin in Ito form}
d\rhopol = \driftK(\rhopol)dt + \sqrt{2\diffusionD(\rhopol)}d\wiener,
\end{equation}
while in \emph{Stratonovich convention}, it becomes
\begin{equation}
\label{eq:Langevin in Stratonovich form}
d\rhopol = \left[\driftK(\rhopol)-\frac{1}{2\jacobian}\dd{}{\rhopol}\jacobian \diffusionD(\rhopol)\right]dt + \sqrt{2\diffusionD(\rhopol)}\circ d\wiener.
\end{equation}
The difference between these two conventions is in the interpretation of how the standard Wiener process, $\wiener$, is integrated. In the \Ito{} calculus, a definite integral of function $g(t)$ is defined as
\begin{equation}
\int\limits_0^T g(t)d\wiener = \lim_{n\rightarrow\infty}\sum\limits_{i=1}^n g_{t_{i-1}}(\wiener_{t_i}-\wiener_{t_{i-1}}),
\end{equation}
whereas Stratonovich definition is
\begin{align}
\int\limits_0^T &g(t)\circ d\wiener = \nonumber\\
&\lim_{n\rightarrow\infty}\sum\limits_{i=1}^n \frac{g_{t_{i}}+g_{t_{i-1}}}{2} (\wiener_{t_i}-\wiener_{t_{i-1}}),
\end{align}
and $0 = t_0 < t_1 < \ldots < t_n = T$. The different definitions lead to practical differences, such that chain rule can be used with Stratonovich but not with \Ito{} convention. Nevertheless the solutions of both formulations of the Langevin equation are equivalent. For more information on these conventions and SDEs in general, see Ref.~\cite{coffey2012langevin}. Here we will find the \Ito{} form, Eq.~\eqref{eq:Langevin in Ito form}, useful in numerical implementation, as it can be solved with explicit methods such as the Euler-Maruyama method. On the other hand, Stratonovich form Eq.~\eqref{eq:Langevin in Stratonovich form} helps us in interpreting the results: The deterministic term in Eq.~\eqref{eq:Langevin in Stratonovich form} has the so-called \emph{noise-induced drift}, $\noisedrift = (1/2\jacobian)(\partial \jacobian\diffusionD / \partial\rhopol)$, separated so that the remnant term, $\nu \equiv \driftK-\noisedrift$, can be extracted from $\driftK$.

A realistic toroidal magnetic geometry typically consists of a core region with well confined field lines together with a stochastic edge region where the field lines are open and ultimately lead particles to vessel wall. The confined core and open edge boundaries are accounted for with reflecting and absorbing boundary conditions, respectively. The stochastic region may contain magnetic islands, and any particles born inside those we assume to be confined indefinitely in the absence of collisions and electric field. To incorporate these field features into our model, we separate confined and unconfined REs,
\begin{equation}
\label{eq:confined+unconfined}
f = f_\mathrm{conf} + f_\mathrm{unconf},
\end{equation}
where $\partial f_\mathrm{conf}/\partial t = 0$. That is, $f_\mathrm{conf}$ consists of particles born within islands or on healthy flux surfaces not experiencing radial transport. Assuming that field lines become stochastic at $\rhopol = \rho_\mathrm{core}$, and that $\rho_\mathrm{edge}$ is the location beyond which all REs are lost practically instantly, we can write the \emph{1D-model} as an initial value problem with boundary conditions
\begin{align}
\label{eq:1Dmodel}
\dd{\tilde{f}}{t} = -\frac{1}{\jacobian}\dd{}{\rhopol}\left[\jacobian\driftK \tilde{f}\right]
&+\frac{1}{\jacobian}\dd{^2}{\rhopol^2}\left[\jacobian\diffusionD \tilde{f}\right],\\
\label{eq:1Dmodel reflecting BC}
\left.\dd{\tilde{f}}{\rhopol}\right\vert_{\rhopol = \rho_\mathrm{core}} &= 0,\\
\label{eq:1Dmodel absorbing BC}
\left.\tilde{f}\right\vert_{\rhopol = \rho_\mathrm{edge}} &= 0,
\end{align}
where $\tilde{f} \equiv f_\mathrm{unconf}$ is used for brevity. It is worth noting that $\rho_\mathrm{edge}$ does not necessarily correspond to the separatrix as the confinement volume shrinks when a particle has a non-negligible orbit width. Close to the $\rho_\mathrm{edge}$ boundary some, but not necessarily all, REs are lost before their orbits can be described as chaotic e.g. during the first poloidal orbit. Obviously, this is not a diffusive process and our model fails to account for it. Nevertheless, we use the 1D-model even there until a more suitable model is devised to account for this laminar transport.

The estimation of the transport coefficients $\driftK$ and $\diffusionD$ with an orbit-following code will be described in the next section. The rest of this section we discuss how 1D-model can be solved using a Monte Carlo method so that it can be benchmarked against the (real) time-evolution obtained with an orbit-following code. There are other means to solve Eqs.~\eqref{eq:1Dmodel}~--~\eqref{eq:1Dmodel absorbing BC} such as the finite element method. The results in this paper were obtained with the Monte carlo method but finite element method was confirmed to yield same results. In the Monte Carlo method, $\tilde{f}$ can be solved by tracing a sufficiently large number of markers according to the equations
\begin{align}
\Delta\rhopol &= K(\rhopol^{i})\Delta t + \beta \sqrt{2D(\rhopol^{i})\Delta t},\\
\rhopol^{i+1} &= \rhopol^{i}+\Delta\rhopol,\\
t^{i+1} &= t^{i} + \Delta t.
\end{align}
Here $\beta \sim \mathcal{N}(0,1)$ is a normally distributed random variable. Time step $\Delta t$ was assigned a constant value of $1\times10^{-8}$~s which is roughly of the order of one circulation time. Markers going beyond $\rhopol=\rho_\mathrm{edge}$ are considered lost and their simulation is terminated. The reflecting boundary condition is satisfied by setting $\rhopol^{i+1} = 2\rho_\mathrm{core} - \rhopol^i-\Delta \rhopol$ for markers that would otherwise have $\rhopol^{i+1} < \rho_\mathrm{core}$. The complete distribution function, Eq.~\eqref{eq:confined+unconfined}, is obtained by flagging each marker as confined or unconfined when the population is initialized. The probability of flagging a marker confined is the ratio between the number of confined markers and the total number of markers born on that location $f_\mathrm{conf}(\rhopol)/f(\rhopol,0)$ which is obtained during the process of evaluating the transport coefficients.

\section{Evaluation of transport coefficients}
\label{sec:Evaluation of transport coefficients}

\subsection{Methods}

The orbit-following code ASCOT~\cite{ascot4ref} is used to find the transport coefficients as well as to verify the 1D-model later on. ASCOT is capable of tracing particle orbits in 3D magnetic fields that contain arbitrary perturbations~\cite{akaslompolo2015iter}. The markers representing REs are traced according to the Hamiltonian guiding center equations of motion~\cite{brizard2007} in cylindrical coordinates $(R,z,\phi)$ in $(\pitch,\energy)$ velocity space. The radial coordinate $\rhopol$ is only used when the marker state is stored.

\begin{figure}[!t]
\centering
\begin{overpic}[width=0.45\textwidth]{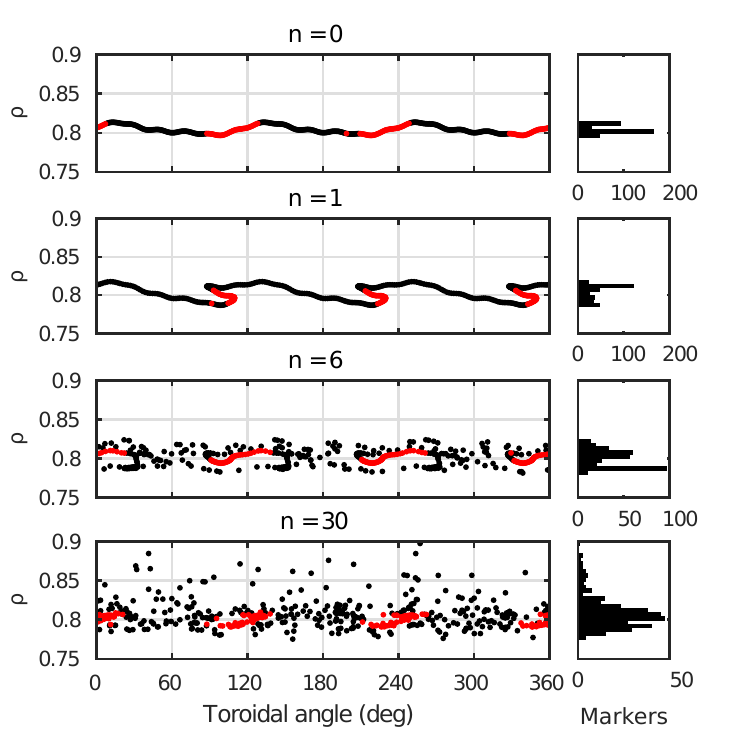}
\end{overpic}
\caption{Evolution of the marker distribution at OMP (left) and the corresponding radial profile (right) of one of the marker populations used in evaluating the coefficients in the background shown in Fig.~\ref{fig:poincare9MA}. Red color indicates markers that are confined in magnetic islands. The radial profile includes only the unconfined markers. $n$ counts the number of poloidal turns completed, with $n=0$ corresponding to the initial configuration. The population is highly correlated at $n=1$ and some correlation remains even at $n=6$. After $n=30$ orbits, the unconfined markers have become uncorrelated and the radial profile resembles that of a (slightly distorted) Gaussian.}
\label{fig:scatter}
\end{figure}

The process of evaluating transport coefficients begins with tracing a population of REs with ASCOT. The population consists of markers with specific energy, pitch, and radial location, which allows one to evaluate corresponding $\driftK(\rhopol,\pitch,\energy)$ and $\diffusionD(\rhopol,\pitch,\energy)$. The markers are populated at the OMP as that location corresponds to the $\rhopol$ coordinate in the 1D-model. In order to capture the stochastic transport correctly, the markers must be distributed toroidally on the chosen $\rhopol$ surface. However, initializing markers in an axisymmetric configuration (since the equilibrium is 2D) would yield noisy transport coefficients since even in the stochastic region remnant flux surfaces exist. These are not axisymmetric as is clearly seen in the bent magnetic islands in Fig.~\ref{fig:poincare9MA}. However, the magnetic perturbation causing the non-axisymmetric flux surfaces is weaker at the high-field side (HFS) so the field structure there is more axisymmetric. We can use this observation to generate an initial population that is better aligned with the remnant flux surfaces at OMP. First, we choose a desired number of axisymmetric (corresponding to the chosen $\rhopol$ value), toroidally uniform points from HFS midplane and then follow field lines from these locations for half a poloidal orbit to the OMP. The marker initialization is now done on the resulting $R,\phi$ positions at OMP, which leads to an initial configuration as the one illustrated in Fig.~\ref{fig:scatter}. Using this initialization method reduces the noise in the evaluated transport coefficients. Note that the $\rhopol$ coordinate we use in recording the marker position is still axisymmetric, but removing this discrepancy would require a more advanced mapping process. In the simulations performed for this paper, we used 500 markers at each radial location, and followed them for 0.1 ms, during which they completed roughly 300 poloidal orbits if not lost before.

The ASCOT simulations provide us a time-evolution of a radial distribution as illustrated in Fig.~\ref{fig:scatter}. Before proceeding to evaluate the transport coefficients, confined markers are removed from the resulting distribution. The confined markers are identified by comparing the trajectories of toroidally adjacent markers to see whether they diverge. Markers lost during the simulation are evidently unconfined. 

There are three ways of evaluating the transport coefficients from the time evolution of the radial distribution in a ideal case where the transport is homogeneous and boundaries extend to infinity. The initially localized marker population, $\tilde{f}(\rhopol_0,0) = \delta(\rhopol_0-\rhopol)$, would evolve as
\begin{equation}
\label{eq:first way}
\tilde{f}(\rhopol,\tau) = \frac{1}{\sqrt{4\pi\diffusionD\tau}}\exp\left[-\frac{(\rhopol-\rhopol_0-\driftK\tau)^2}{4\diffusionD\tau}\right].
\end{equation}
Thus the first method would be to simply fit distribution~\eqref{eq:first way} using $\driftK$ and $\diffusionD$ as parameters. The second method is even more straightforward as it requires no curve fitting: The transport coefficients are evaluated by noting that Eq.~\eqref{eq:first way} is a Gaussian distribution so its mean $\mu_\rhopol$ and variance $\sigma_\rhopol^2$ are given by
\begin{equation}
\label{eq:second way}
\driftK = \frac{\mu_\rhopol-\rhopol_0}{\tau} \quad\textrm{and}\quad \diffusionD = \frac{\sigma_\rhopol^2}{2\tau},
\end{equation}
The third way is to first define a coordinate $\rhoedge$, and then find for each marker the time instant it first crosses $\rhoedge$. The distribution of this first passage time, $T_{\rhoedge}$, obeys the Inverse-Gaussian distribution
\begin{equation}
\label{eq:third way}
T_{\rhoedge}(t) = \sqrt{\frac{c_2}{2\pi t^3}}\exp\left[\frac{-c_2(t-c_1)^2}{2c_1^2t}\right],
\end{equation}
where $\delta\rhopol = \rhoedge-\rhopol_0$, $c_1 = \delta\rhopol/\driftK$ and $c_2 = 2(\delta\rhopol)^2/\diffusionD$, and we have assumed $\delta\rhopol > 0$ and $\driftK > 0$. Again the transport coefficients are obtained with curve fitting.

\begin{figure}[!t]
\centering
\begin{overpic}[width=0.45\textwidth]{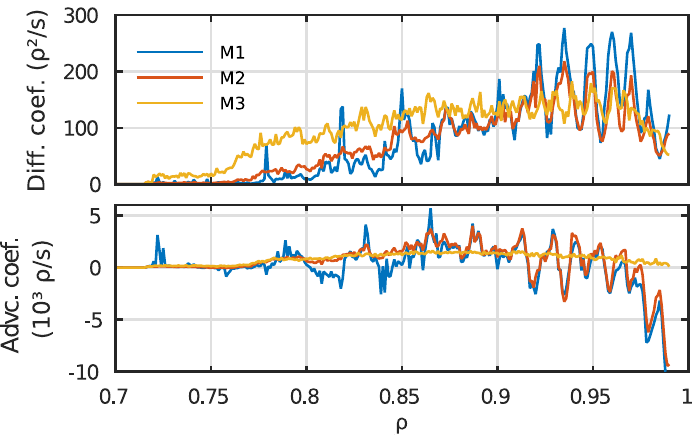}
\end{overpic}
\caption{Comparison of transport coefficients obtained with methods introduced in Eqs.~\eqref{eq:first way}, \eqref{eq:second way}, and \eqref{eq:third way} (M1, M2, and M3 in the legend, respectively.) M1 and M2 overlap for most part but M2 has less noise in its profile. The similarity shows that the profile evolves roughly as Gaussian~\eqref{eq:first way} despite the spatial variation in transport coefficients. M3 contains least oscillations near the edge ($\rhopol > 0.9$),  but its $\diffusionD$ profile deviates significantly from other methods closer to the core. M3 results in coefficients averaged over the region from $\rhopol_0$ to $\rhopol_\mathrm{edge}$ which explains the large differences. To transform the coefficients from units of $\rhopol$ to meters, one needs to multiply $K$ with $\jacobian$ and $D$ with $\jacobian^2$. The methods were compared in the background shown in Fig.~\ref{fig:poincare9MA} for which Jacobian $\jacobian \approx 1.8$ m.}
\label{fig:methods}
\end{figure}

All the listed methods give same results in the ideal case. However, we have to choose the most suitable method since we expect $\driftK$ and $\diffusionD$ to have spatial dependency and wish to calculate them even in the proximity of the boundaries. None of the methods are, strictly speaking, applicable when the coefficients are spatially dependent but they still produce reasonable results as will be shown later. The reflecting boundary near the core and the absorbing boundary at the edge prevents using the method \eqref{eq:first way} close to these as the distribution would be far from Gaussian. The second method \eqref{eq:second way} is easily biased by outliers, i.e. individual markers that have moved to a region with a significantly higher transport, whereas the first is not. For both methods, $\tau$ has to be chosen so that the population has become uncorrelated (recall Fig.~\ref{fig:scatter}) but still short enough so that no markers are lost or the population allowed to spread too wide. Near the edge, the natural choice is Eq.~\eqref{eq:third way} where $\rhoedge =\rhopol_\mathrm{edge}$. Figure~\ref{fig:methods} shows how the methods compare against one another. 

For the 1D-model, the coefficients were evaluated by combining methods~\eqref{eq:second way} and~\eqref{eq:third way}. The latter was used for marker populations that had sufficient losses. Coefficients were not evaluated in the region where the transport is laminar. Instead, the coefficients were extrapolated using a constant value. When the method~\eqref{eq:second way} was used, $\tau$ was chosen to be either the time when the first marker was lost or the simulation time (0.1 ms) if there were no losses. We did not find it necessary to separately account for the population spreading when choosing $\tau$ in this work.

\subsection{Numerical values for the transport coefficients}

\begin{figure}[!t]
\centering
\begin{overpic}[width=0.45\textwidth]{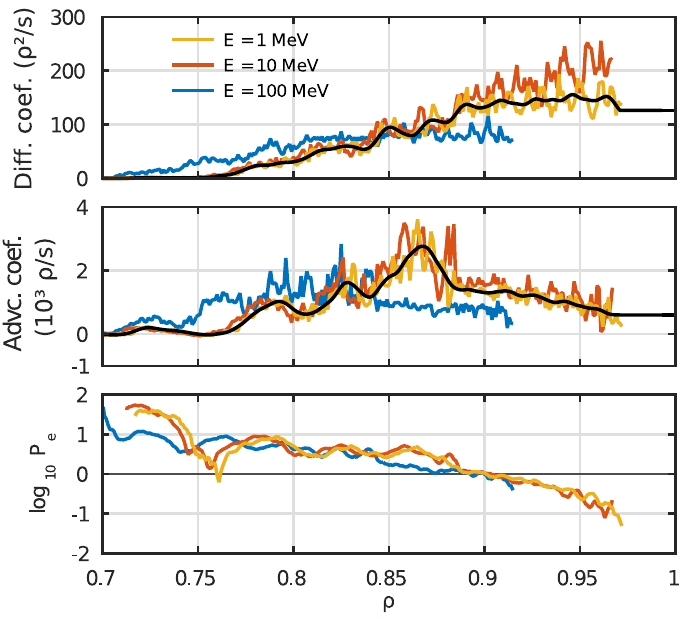}
\put(16,84){a)}
\put(16,57){b)}
\put(16,13){c)}
\end{overpic}
\caption{Diffusion (a) and advection (b) coefficient profiles for REs with different energies ($\pitch = 0.9$ for all) calculated in a field shown in Fig.~\ref{fig:poincare9MA}. The coefficients are cut off from the first flux surface where first-orbit losses exist. In order to reduce the level of noise originating from the evaluation process before applying them in the 1D-model, the coefficients were spatially smoothed with LOESS method with a suitable smoothing parameter. The smoothed coefficients for ($\energy = 1$ MeV, $\pitch = 0.9$) case are shown with black curves. (c) The \Peclet{} number, on a logarithmic scale, indicating whether the advection or the diffusion is the dominant transport mechanism.}
\label{fig:coefficients_energy9MA}
\end{figure}

The transport coefficients for the field introduced in Fig.~\ref{fig:poincare9MA} were calculated for 300 equally spaced radial positions from $\rhopol = 0.7$ to $\rhopol = 1.0$. They are shown in Fig.~\ref{fig:coefficients_energy9MA} for different energies. We now proceed to analyse the small orbit width (the yellow curve) case alone and discuss the relation between pitch, energy and transport later.

The stochastic region begins at $\rhopol = 0.72$ below which the coefficients are zero. The diffusion coefficient is small and starts to grow only after passing the major islands at $\rhopol = 0.75$. The growth is generally linear but there are dips at $\rhopol = 0.80$, $0.84$ and $0.86$ where magnetic islands are located, and one at $0.88$. The profile saturates after the last dip and the remaining oscillations are likely artefacts from the evaluation process. The advection coefficient begins with a positive value but there it is probably an artefact due to the reflecting boundary at $\rhopol = 0.72$. The negative slope leads to a negative advection coefficient at $\rhopol \approx 0.75$. Similar to $\diffusionD$, $\driftK$ grows linearly after that with dips at $\rhopol = 0.80$ and $0.84$. The advection coefficient has a maxima at $0.87$, after which it decreases towards the edge. The dips confirm that the large islands around $\rhopol = 0.80$ and $0.84$ create transport barriers where both advection and diffusion are reduced. At $\rhopol = 0.75$, $\driftK$ is negative while $\diffusionD$ is small which indicates a significantly restrained transport there.

Evidently advection is present in all regions where there is transport but this does not mean it necessarily has an important role. The importance of the advection with respect to diffusion is characterised by the \Peclet{} number, $P_e \equiv L|\driftK|/\diffusionD,$ where $L$ is the characteristic length scale of the transport process. For $P_e \gg 1$ advection dominates over diffusion, while for $P_e \ll 1$ the opposite is true. Choosing $L = \rhopol_\mathrm{edge}-\rhopol_0$, we find (Fig.~\ref{fig:coefficients_energy9MA}~(c)) that $P_e < 1$ from $\rhopol = 0.89$ onwards, i.e., after the last island. The profile of $P_e$ is not monotonous but peaks in island regions. The strong dip at $\rhopol = 0.76$ is due to the sign change in the advection coefficient. This analysis show that advection is either dominant or of equal importance as the diffusion when islands are present. When the field becomes more stochastic, the importance of the advection is reduced.

\subsection{An alternative method to evaluate the diffusion coefficient}
\begin{figure}[!t]
\centering
\begin{overpic}[width=0.45\textwidth]{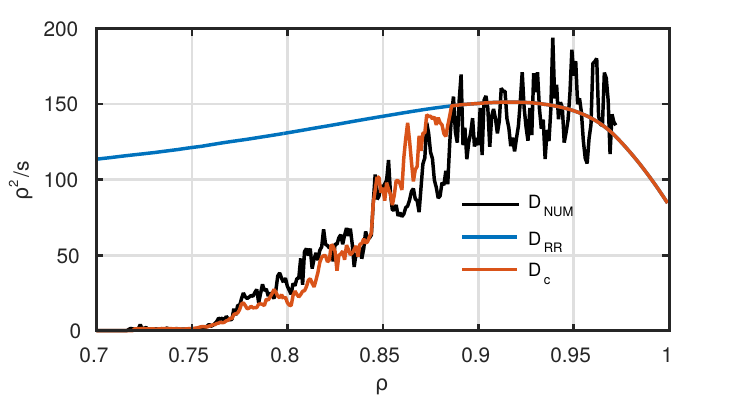}
\end{overpic}
\caption{Numerical diffusion coefficient $\Dnum$ for $(E = 1$  MeV, $\xi = 0.9)$ electrons compared to the Rechester Rosenbluth result $\DRR$, Eq.~\eqref{eq:RR diffusion coeff}, and to the form $\Dc$, Eq.~\eqref{eq:fitted diffusion coeff}. The coefficients are for the field in Fig.~\ref{fig:poincare9MA}. Here, the constant $C$ has a value of $C = 0.025$ for both $\DRR$ and $\Dc$. Note that $\DRR$ and $\Dc$ overlap when $\rhopol > 0.88$.}
\label{fig:DfitAndCorrLength}
\end{figure}

Figure~\ref{fig:DfitAndCorrLength} compares the numerical values for $\diffusionD$ to the Rechester-Rosenbluth result, Eq.~\eqref{eq:RR diffusion coeff}. $\DRR$ overestimates transport significantly outside the edge region and should not be used in regions where the remnant islands exist. However, $\DRR$ captures the overall trend at the edge $(\rhopol > 0.9)$ where islands are not present (recall Fig.~\ref{fig:coefficients_energy9MA}).

However, we found that diffusion coefficient of form
\begin{equation}
\label{eq:fitted diffusion coeff}
\Dc \equiv C\frac{\sigma^2}{2\tc }\tilde{b}^2,
\end{equation}
which one can obtain by replacing the orbit circulation time in Eq.~\eqref{eq:RR diffusion coeff} with a correlation time $\tc$, agrees with the Monte-Carlo result. The form~\eqref{eq:fitted diffusion coeff} is based on the observation that initially closely located test particle orbits tend to be correlated for a time $\tc$ before the transport of the particle population can be described as diffusive (as was seen in Fig.~\ref{fig:scatter}). A similar observation of clumping in a stochastic field was first made in Ref.~\cite{eijnden1995liouvillian}, where two closely located field lines were found to be correlated for a time before the mean-square relative distance between the field lines started to grow exponentially. This process is evident where magnetic islands are present and the correlation time can be several times larger than the orbit circulation time $(\tc\gg\tau_{orb})$ which leads to a significantly reduced diffusion coefficient as indicated in Fig.~\ref{fig:DfitAndCorrLength}. The proposed form, $\Dc$, is more suitable in the inner region where $\DRR$ fails. Close to the edge, the correlation time becomes negligible as $\tc < \tau_\mathrm{orb}$, and Eq.~\eqref{eq:fitted diffusion coeff} reduces to Eq.~\eqref{eq:RR diffusion coeff}. We observe that this occurs after the last island surface. $\tc$ in Eq.~\eqref{eq:fitted diffusion coeff} was evaluated by particle tracing as described in Appendix A. Although this does not yet provide a formulation from which diffusion coefficient could be evaluated directly from the magnetic field properties without tracing particles, Eq.~\eqref{eq:fitted diffusion coeff} still allows for independent evaluation of $\diffusionD$. 

\section{Simulations with 1D-model}
\label{sec:Benchmark to ASCOT solution}

\subsection{Comparison of 1D-model to full 3D simulations}

\begin{figure}[!t]
\centering
\begin{overpic}[width=0.45\textwidth]{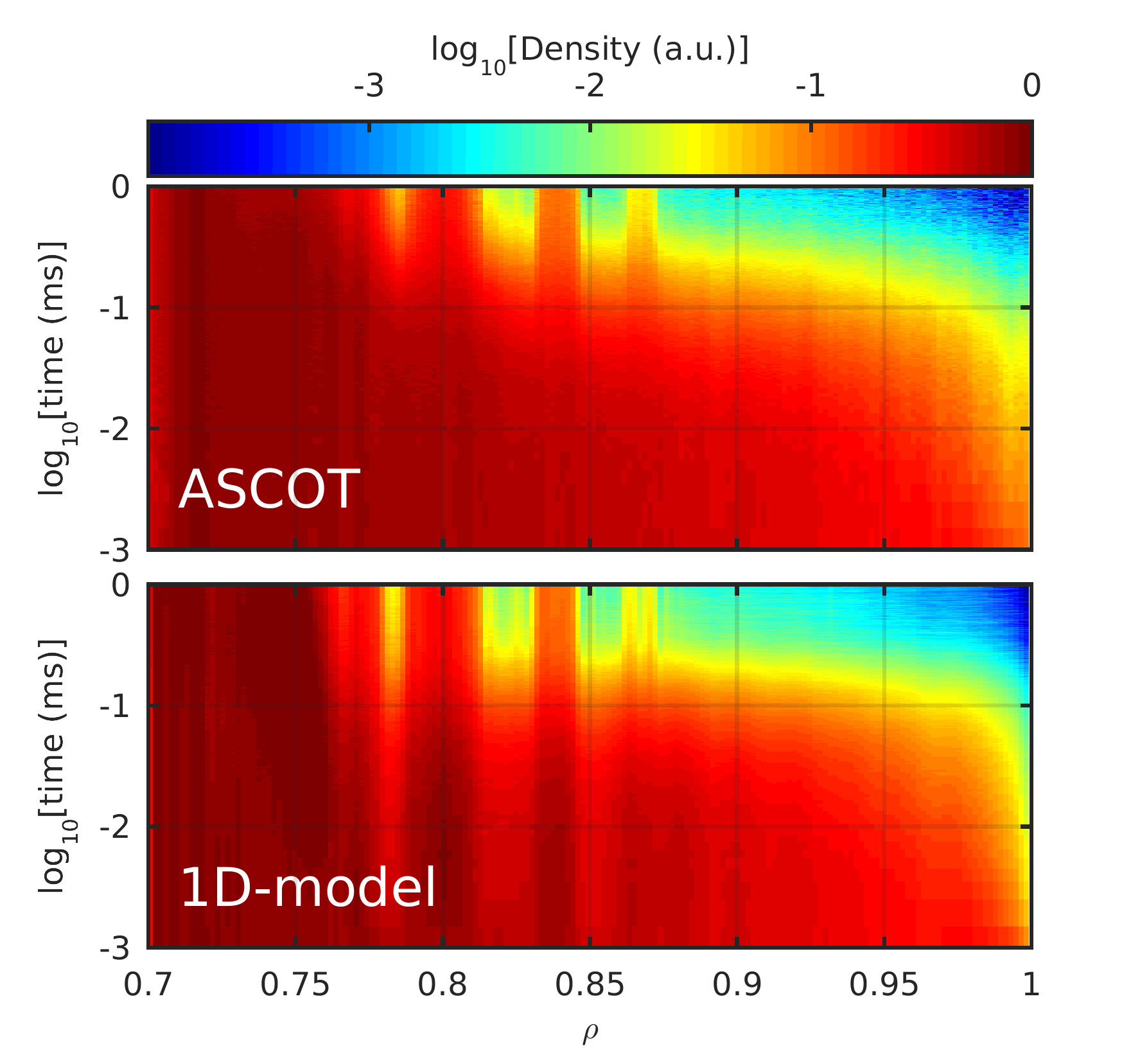}
\put(15,72){\textcolor{white}{a)}}
\put(15,36){\textcolor{white}{b)}}
\end{overpic}
\caption{The particle density evolution in ASCOT (a) and 1D-model (b) for electron population with ($\energy = 1$ MeV, $\pitch = 0.9$) in a magnetic field shown in Fig.~\ref{fig:poincare9MA}. The full 3D simulation with ASCOT required approximately 600 cpu-hours while only a few minutes were needed with the 1D-model.}
\label{fig:rhoTimeEvolutionRE_E6_p0.9_9MA}
\end{figure}

The 1D-model was benchmarked by carrying out equivalent simulations with both 1D-model and ASCOT in the background shown in Fig.~\ref{fig:poincare9MA}. The edge region was populated uniformly in $\rhopol$ (the mapping process described in Section 3 was also used here) and markers were traced for 1 ms, during which almost all unconfined markers were lost. The population consisted of electrons with given $\pitch$ and $\energy$. Here we will present the results for the field whose structure was shown in Fig.~\ref{fig:poincare9MA}, and for the ($\energy = 1$ MeV, $\pitch = 0.9$) case only. The smoothed coefficients used in the 1D-model, \eqref{eq:1Dmodel}, were already shown in Fig.~\ref{fig:coefficients_energy9MA}. The results of the benchmark simulations are show in Fig.~\ref{fig:rhoTimeEvolutionRE_E6_p0.9_9MA} which we now proceed to discuss.

As the ASCOT result is from an orbit-following simulation, it can be regarded as a correct picture of how the RE density evolves. The evolution of the density is smooth from $\rhopol = 0.87$ to the very edge. In the inner region, this smoothness disappears when protrusions, caused by the markers confined to the magnetic islands, appear. The magnetic islands also cause a step-like structure, most easily seen by following the contrast between the darker and lighter red, which indicates presence of transport barriers at locations $\rhopol = 0.76$, $0.82$, and $0.86$. Noticeably, the density around $\rhopol = 0.75$ stays constant all the way until 0.5 ms has passed, meaning that the transport is severely reduced by the barrier at $\rhopol = 0.76$.

The 1D-model is able to replicate ASCOT results to a good extent. The agreement is good in the region with no islands ($\rhopol > 0.86$) with the exception of the very edge where the 1D-model represents transport as a diffusive process when in fact it is laminar. At the inner region, the 1D-model density profile is slightly more filamental as the markers have clustered around the islands. Despite this clustering, the average time it takes for a marker starting from a given position to be lost was found to be approximately the same for ASCOT and 1D-model in all regions. The only exception is the region $\rhopol < 0.76$, where in the 1D-model density increases in time. There the 1D-model overestimates the transport barrier, hindering markers from escaping. Eventually, even those markers will escape but at a rate lower that in the ASCOT simulation.

The benchmark was repeated for several values of pitch and energy, and also for different magnetic configuration. The results are collected in Appendix B.  Both fields were reconstructed using the same methods as those used in ITER fast ion simulations~\cite{akaslompoloFDE2015}. 
Even though both magnetic configurations used in the benchmarks were reconstructed using the same methods as those used in ITER fast ion simulations~\cite{akaslompoloFDE2015}, one should note that neither field represent plasma state after a disruption. Therefore, no conclusions should be made on ITER RE-mitigation based on these results. The sole purpose here is to test and develop the 1D-model, and for this these fields are applicable. 

\subsection{Advection coefficient analysis}

\begin{figure}[!ht]
\centering
\begin{overpic}[width=0.45\textwidth]{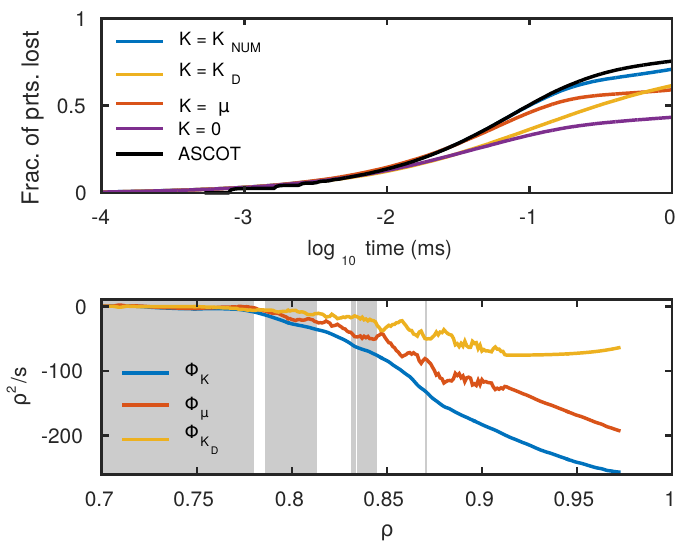}
\put(6,80){a)}
\put(6,39){b)}
\end{overpic}
\caption{Analysis of the advection coefficient and its components. (a) Fraction of particles lost as a function of time for simulations where advection coefficient was modified. The loss rate evaluated with ASCOT has a step-like structure (around $t=1$ $\mu$s) which is caused by the laminar losses. (b) Potentials associated with the advection terms. The potentials were evaluated using the form Eq.~\eqref{eq:fitted diffusion coeff} for $\diffusionD$ to avoid the edge oscillations present in $\Dnum$. Grey regions indicate magnetic island locations in Fig.~\ref{fig:poincare9MA}. The coefficients in both plots correspond to ($\energy = 1$ MeV, $\pitch = 0.9$) case.}
\label{fig:driftAnalysis}
\end{figure}

The \Peclet{} number already indicated that the advection plays an important role especially in the presence of the magnetic islands (recall Fig.~\ref{fig:coefficients_energy9MA}). Here we continue this discussion by analysing the different components of the advection. The 1D-model simulation was repeated in different cases where the numerically evaluated advection coefficient, $\driftK_\mathrm{NUM}$, was replaced by the noise induced drift $\noisedrift$, the remnant advection $\nu = \driftK_\mathrm{NUM} - \noisedrift$, or was set to zero. Comparing the loss rates from these simulations, shown in Fig.~\ref{fig:driftAnalysis}~(a), shows how each subsequent alteration leads to less representative results, with zero-advection being the least accurate. $\driftK = \nu$ case was closest to the original but losses saturated more quickly, meaning some transport barriers were enhanced when noise-induced drift was omitted. However, the results with noise-induced drift alone lags behind the ASCOT results already after 0.01 ms, indicating that the remnant component $\nu$ forms an essential part of the advection.

For the advection coefficient we could not find a similar formulation as was found for the diffusion coefficient (Eq.~\eqref{eq:fitted diffusion coeff}). Some analysis can still be made which could reveal more about the nature of the remnant advection. The advection coefficient may be thought of as a pseudo-force and, as it depends only on $\rhopol$ making it conservative, the related potential, $\Phi_K$, can be found via relation $K = -\nabla \Phi_K$. Likewise, we have $\Phi_\nu$, for the remnant advection and $\Phi_{\noisedrift} = \diffusionD/2$ for the noise-induced drift. The potentials, shown in Fig.~\ref{fig:driftAnalysis}~(b), have global minima at the edge except for $\Phi_{\noisedrift}$ which saturates before that. $\Phi_\nu$ decreases towards the edge while it also has a local minima at islands. This shows that $\nu$ has a twofold role: it models the effect of islands as attractors, capturing transport barriers, while it also models the non-diffusive flow of particles from the plasma when the field is completely stochastic. The minima around the islands are partly countered by decreasing $\Phi_{\noisedrift}$ which explains the behaviour seen in Fig.~\ref{fig:driftAnalysis}~(a) when the advection coefficient was altered. 

\section{Pitch and energy dependency of transport}
\label{sec:Pitch and energy dependency of transport}

We have shown that 1D-model is well-suited for modelling the radial transport but there remains one issue to be dealt with before it can be coupled to an orbit-averaged tool. Applying the 1D-model to a real RE population that does not have a discreet pitch and energy values requires a robust interpolation scheme for the transport coefficients. This in turn requires knowledge of how the transport depends on RE parameters. 

\begin{figure}[!t]
\centering
\begin{overpic}[width=0.45\textwidth]{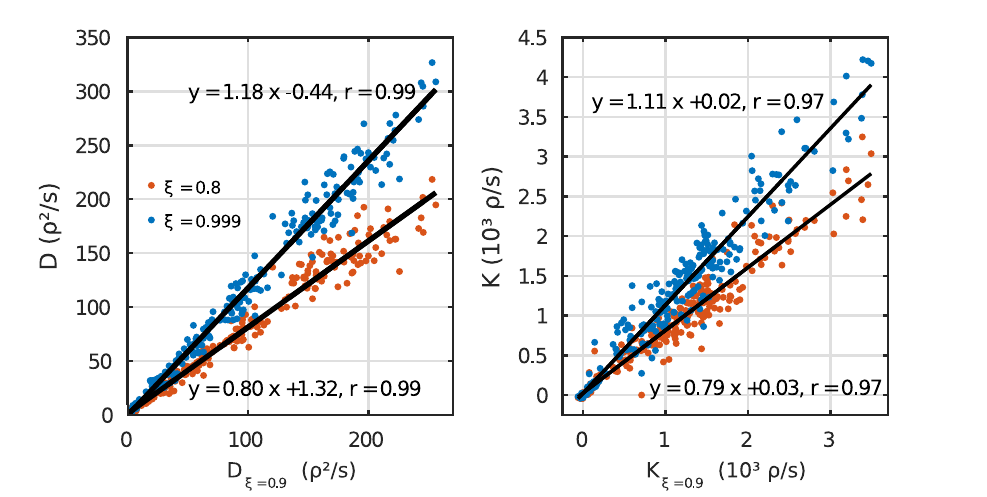}
\put(14,43){a)}
\put(58,43){b)}
\end{overpic}
\caption{Pitch dependency on transport coefficients. The coefficients were obtained in a manner similar to Fig. 4, except that in this case pitch was varied. (a) Diffusion coefficients for REs ($\energy = 10$ MeV) with $\pitch = 0.8$ (orange dots) and $0.999$ (blue dots) plotted versus $\diffusionD$ for $\pitch = 0.9$ at different radial positions. Black lines are linear fits with $r$ being the correlation coefficient. (b) Same for the advection coefficient. The coefficients correspond to the magnetic configuration in Fig.~\ref{fig:poincare9MA}. }
\label{fig:coefficients_pitch9MA}
\end{figure}

To find out how the transport depends on the pitch value, we evaluated the coefficients for populations that were identical except for the pitch. We chose large pitch values, $\pitch = 0.8$, $\pitch = 0.9$, and $\pitch = 0.999$, to reflect the fact that REs have a pitch close to unity. The results, collected in Fig.~\ref{fig:coefficients_pitch9MA}, indicate that linear interpolation is sufficient. The caveat is that the slope in the linear fits is not the pitch ratio itself which is what Eq.~\ref{eq:RR diffusion coeff} would have suggested: With the higher pitch value ($\pitch = 0.999$), the dependency is closer to the expected one as if the transport was solely due to the particles following chaotic field lines. This is no longer the case with the lower pitch value as the orbit width becomes larger. In conclusion, linear interpolation is feasible but the non-universal multiplication factor requires that transport coefficients are tabulated for sufficiently many pitch values.

\begin{figure}[!t]
\centering
\begin{overpic}[width=0.45\textwidth]{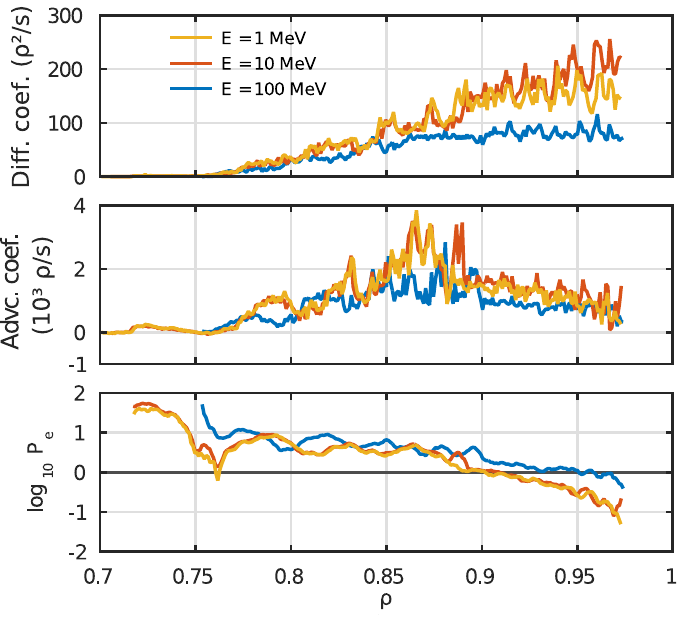}
\put(16,85){a)}
\put(16,57){b)}
\put(16,13){c)}
\end{overpic}
\caption{The transport coefficients and \Peclet{} number from Fig.~\ref{fig:coefficients_energy9MA} shifted radially outwards in $\rhopol$ by a $\Delta\rhopol = [\textrm{orbit width}]/\sqrt{2}$, which is the RMS value of the particle excursion from its initial $\rhopol$ surface over the course of its poloidal turn.}
\label{fig:coefficients_energycorrected9MA}
\end{figure}

The change in transport with respect to $\energy$ is more essential to understand as energy can vary by orders of magnitude within a RE population. The large differences in Fig.~\ref{fig:coefficients_energy9MA} between different energies cannot be explained by change in $v_\parallel$ since $v_\parallel \approx 0.94 c$ already at 1 MeV. However, increase in particle energy also increases the orbit width, and the particle sweeps radially wider region during one poloidal orbit. Therefore, a particle is exposed to field structures outside its initial surface $\rhopol_0$, to regions where transport can be significantly different in magnitude. It turns out that understanding this effect helps us in performing the energy interpolation. In Fig.~\ref{fig:coefficients_energycorrected9MA} we have replotted the results of Fig.~\ref{fig:coefficients_energy9MA} by moving the $\rhopol$-label of the coefficients calculated for a given surface outward by a step $\Delta\rhopol = [\textrm{orbit width}]/\sqrt{2}$. The shifted coefficients between different energies have more overlap between them. However, the coefficients for $\energy = 10$ MeV and $\energy = 100$ MeV still have qualitatively different profiles, and coefficients for e.g. $\energy = 50$ MeV cannot be interpolated with certainty. Additional coefficient evaluations with different $\energy$ are therefore called for, but the shifting mechanism reduces the number of evaluations required.

We can now use the results we have gathered to discuss the energy-transport relation. Comparing the transport of REs with different energies using the coefficients in Fig.~\ref{fig:coefficients_energy9MA} is questionable as energy also affects the confinement volume. For example at $\rhopol = 0.9$, electrons with $\energy = 100$ MeV have smaller values for both coefficients  when compared to the ones with $\energy = 10$ MeV. Despite this, simulations reveal that the electrons with higher energy are lost at higher rate when launched from this surface. This is simply because they have less distance to cover before they are lost due to their large orbit width. On the other hand, the shifted coefficients in Fig.~\ref{fig:coefficients_energycorrected9MA} all have their cut-off points (beyond which markers are lost within a few orbits) approximately at the same $\rhopol$ coordinate. This means that $\Delta\rhopol$ equals to the confinement volume shrinkage, and, with Fig.~\ref{fig:coefficients_energycorrected9MA}, the transport of electrons with different energies can be compared on a basis where the distance to the absorbing boundary is the same. Now we can make an interesting observation: Despite the reduction of diffusion coefficient, which was already observed in Ref.~\cite{myra1993quasilinear}, the overall transport of more energetic REs is not necessarily reduced to a same extent. The transport could be increased as the (positive) advection coefficient is more dominant in $\energy =$ 100 MeV case. This aspect should be studied more in relevant backgrounds as it might affect how the RE mitigation should be planned.

\section{Conclusions}
\label{sec:Conclusions}

We have shown that runaway electron radial transport in a perturbed magnetic field can be modelled with an advection-diffusion model with good accuracy. The transport coefficients capturing the effect of the 3D magnetic field have been evaluated with an orbit-following code. However, the spatial dependency of these coefficients, the boundaries of the stochastic region, as well as the particle clumping, complicate the evaluation process. 

As expected, the magnetic islands have a significant role in RE transport. The islands were found to both confine particles born inside them as well as attract particles travelling in the stochastic field. Although this has been recognized before, we have now provided a quantitative measure for the latter effect in the form of an advection coefficient. As for the diffusion coefficient, the Rechester-Rosenbluth result was found to provide, in the zero orbit width limit, a good approximation in a completely stochastic region but it should not be used when magnetic islands are present. We showed that including the orbit correlation time significantly improved the estimate for the diffusion coefficient.

Future work involves incorporating the developed model into an orbit-averaged code, to account for transport induced by magnetic perturbations. The model itself could be improved by addressing the laminar flow at the edge boundary with a suitable sink term. Other important topics are the origin of the advection term and the exact relationship between runaway energy and transport.

\ack

This work is part of the EUROFUSION Enabling Research project ER15-CEA-09. The work was partially funded by Fusion For Energy Grant 379 and the Academy of Finland project No. 259675, and has also received funding from Tekes – the Finnish Funding Agency for Innovation under the FinnFusion Consortium. The simulations performed for this work were carried out using the computer resources within the Aalto University School of Science `Science-IT' project. 

\appendix

\section{Evaluation of $\tc$}

Equation~\eqref{eq:fitted diffusion coeff} requires the evaluation of the correlation time $\tc$ which is described here. $\tc$ characterizes for how long particles launched from close initial positions trace almost similar trajectories before they start to diverge. Confined particles have $\tc = \infty$ as they are not transported by the stochastic field, and thus can be excluded. When the particle orbits are chaotic, the separation of two initially closed particles grows as
\begin{equation}
\label{eq:separation exp}
|\separation(t)| \sim |\separation(0)|\exp(\lambda t),
\end{equation}
where $|\separation(0)|$ is the initial distance between the particles. Here $\lambda$ is the \emph{Lyapunov exponent} which characterizes the chaos of the system (for chaotic orbits $\lambda > 0$). The characteristic time-scale of the exponential growth for the whole population, $\tcorr$, can be estimated as an inverse of the \emph{maximal Lyapunov exponent}~\cite{abdullaev2014magnetic}, defined as
\begin{equation}
\label{eq:maximal Lyapunov exponent}
\tcorr^{-1} = \lambda \equiv \lim_{t\rightarrow\infty}\lim_{\left| \separation(0) \right| \rightarrow 0}\frac{1}{t}\ln \frac{\left|\separation(t)\right|}{\left|\separation(0)\right|}.
\end{equation}
The separation~\eqref{eq:separation exp} does not feature the linear part during which markers are correlated. However, it turns out that $\tc$ is proportional to $\tcorr$~\cite{eijnden1995liouvillian}, and $\tc$ can be solved from Eq.~\eqref{eq:maximal Lyapunov exponent}.

We estimated $\lambda$ numerically by first evaluating the separation in $\rhopol$ coordinate for each adjacent pair of markers initialized at the same flux surface till the end of the simulation time, or until other marker was lost. The same unconfined markers were used as those used in evaluating the coefficients. The Lyapunov exponent for each pair was calculated as
\begin{equation}
\lambda_i = \frac{1}{T_i-t_i}\ln \frac{|\Delta\rhopol_i(T_i)|}{|\Delta\rhopol_i(t_i)|},
\end{equation}
where $T_i$ is the final time instant, and $\Delta\rhopol_i(t_i)$ is the separation after the first completed poloidal orbit. For each populated $\rhopol$ surface, we then estimated the correlation time as $\tau_K(\rhopol) = 1/\max\{\lambda_i(\rhopol)\}$. Note that even though the same markers were used in evaluating both $\Dnum$ and $\lambda$, taking the maxima in Eq.~\eqref{eq:maximal Lyapunov exponent} ensures that Eq.~\eqref{eq:fitted diffusion coeff} is not connected to $\Dnum$ by the numerical evaluation process.

\section{Additional benchmark results}

\begin{figure}[!ht]
\centering
\begin{overpic}[width=0.45\textwidth]{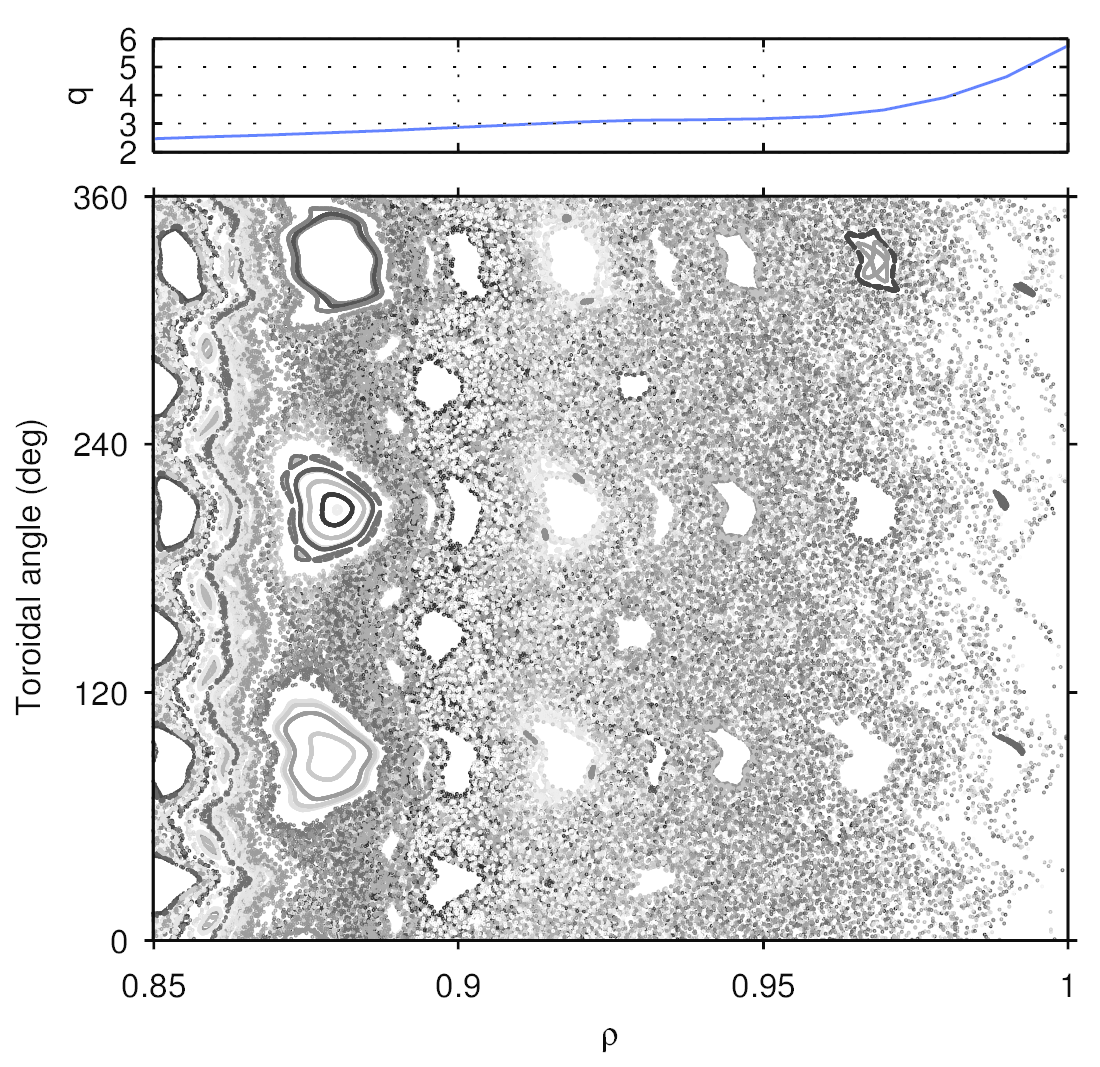}
\end{overpic}
\caption{The magnetic field structure corresponding to ITER $I_p = 15$ MA baseline scenario involving perturbations from ferritic inserts, test blanket modules, and RMP coils. The RMP coil current configuration is: mode $n=3$ and amplitude 45 kAt maximum current. Here $\jacobian = \partial R/\partial\rhopol \approx 2.2$ m.}
\label{fig:poincare15MA}
\end{figure}

\begin{figure}[!ht]
\centering
\begin{overpic}[width=0.45\textwidth]{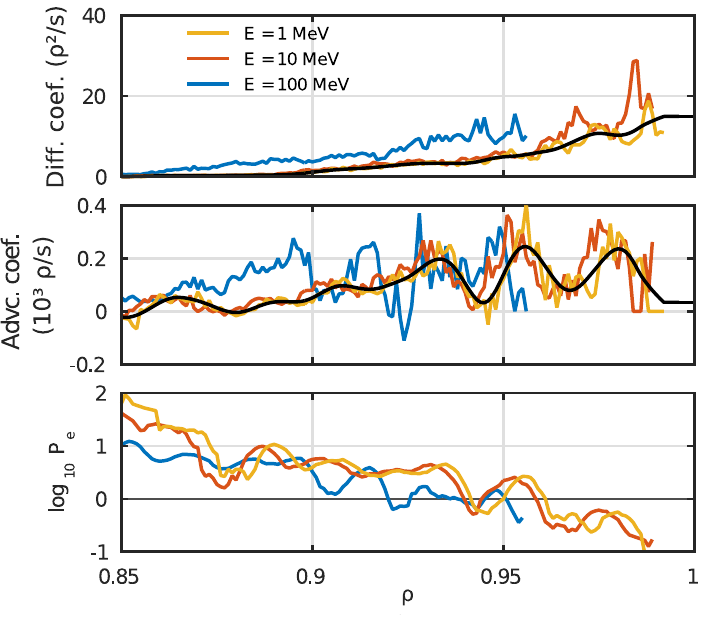}
\put(18,82){a)}
\put(18,55){b)}
\put(18,13){c)}
\end{overpic}
\caption{The transport coefficients in the magnetic background shown in Fig.~\ref{fig:poincare15MA}. (a) Diffusion (b) and advection coefficient, and (c) \Peclet{} number for REs with different energies ($\pitch = 0.9$ for all).}
\label{fig:coefficients_energy15MA}
\end{figure}

\begin{figure}[!ht]
\centering
\begin{overpic}[width=0.45\textwidth]{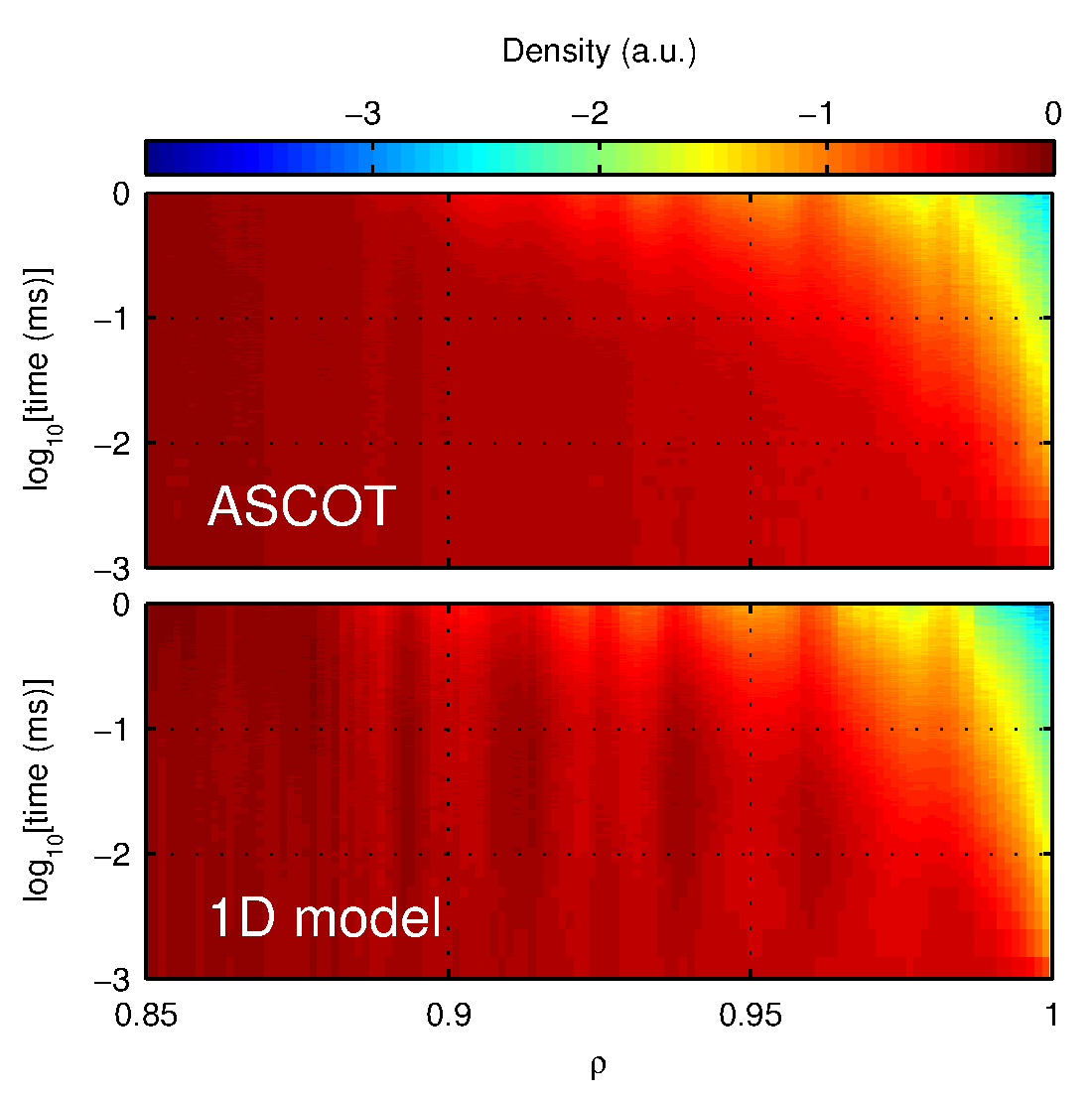}
\put(15,72){\textcolor{white}{a)}}
\put(15,37){\textcolor{white}{b)}}
\end{overpic}
\caption{Evolution of particle density in ASCOT and 1D-model for electron population with ($\energy = 1$ MeV, $\pitch = 0.9$) in the magnetic background shown in Fig.~\ref{fig:poincare15MA}.}
\label{fig:rhoTimeEvolutionRE_E6_p0.9_15MA}
\end{figure}

The transport studies were replicated using a different background shown in Fig.~\ref{fig:poincare15MA}. The structure of this field differs significantly from the one used earlier (recall Fig.~\ref{fig:poincare9MA}). A qualitative inspection suggests that the transport is lower in this case as the field stochasticity seems to be lower and the remanent magnetic islands cover wider regions. Transport coefficients for different energies were evaluated at 150 equally spaced positions between $\rhopol = 0.85$ and $\rhopol = 1.0$, and the results are collected in Fig.~\ref{fig:coefficients_energy15MA}. The results confirm the reduction in the transport as both advection and diffusion coefficient are roughly an order of magnitude less compared to the earlier case. However, the 1D-model is still able to replicate the transport fairly well even in this background as shown in Fig.~\ref{fig:rhoTimeEvolutionRE_E6_p0.9_15MA}.

\begin{figure}[!ht]
\centering
\begin{overpic}[width=0.45\textwidth]{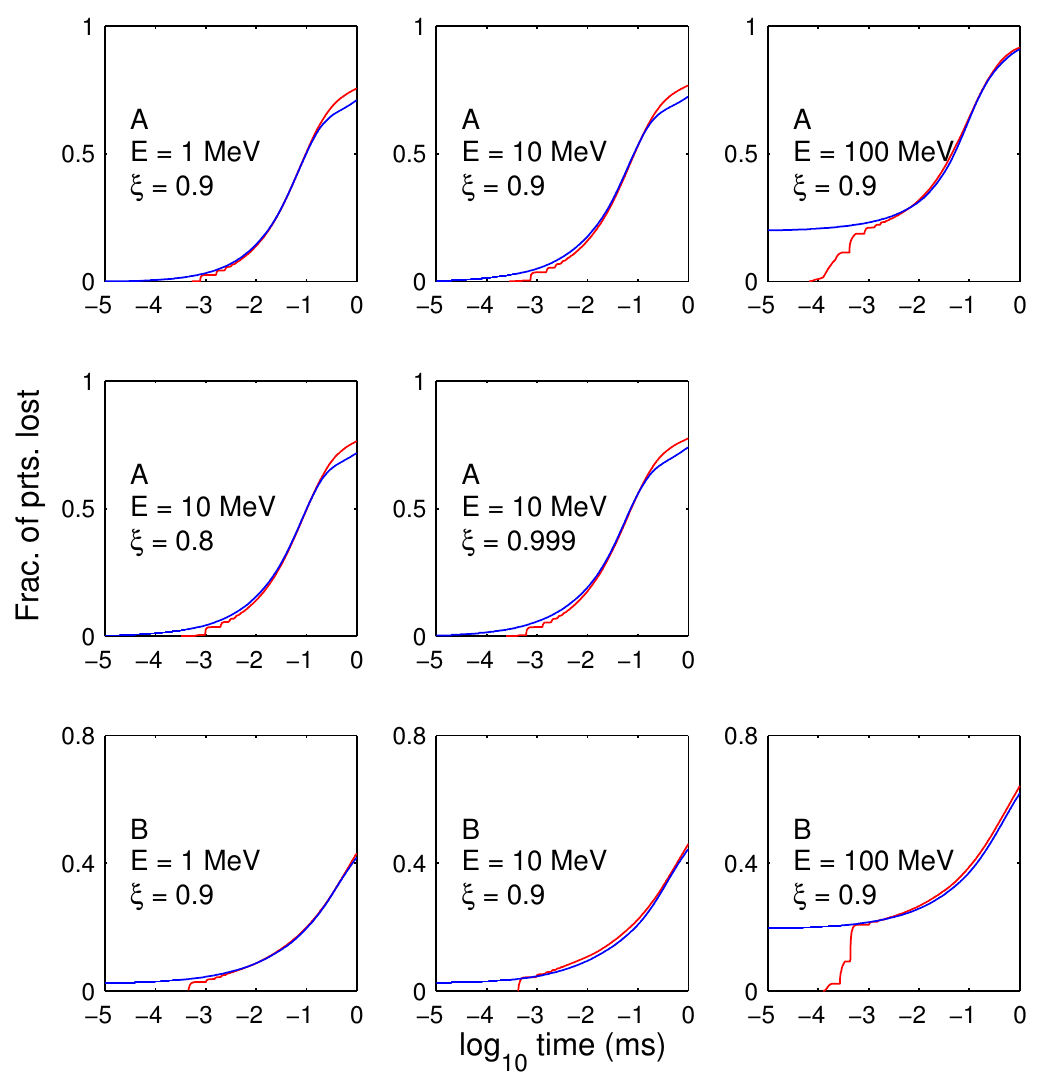}
\put(10.5,93){a)}
\put(41,93){b)}
\put(72,93){c)}
\put(10.5,60){d)}
\put(41,60){e)}
\put(10.5,27.5){f)}
\put(41,27.5){g)}
\put(72,27.5){h)}
\end{overpic}
\caption{Comparison of average loss rates calculated with 1D-model (blue curves) and ASCOT (red) in different cases. Each plot is marked with test particle energy and pitch, and letter \textbf{A} refers to the magnetic background shown in Fig.~\ref{fig:poincare9MA} while \textbf{B} refers to the one in Fig.~\ref{fig:poincare15MA}. The step-like structure in ASCOT loss rate and corresponding instant losses in 1D-model are due to the laminar transport.}
\label{fig:allLossrates}
\end{figure}

The comparison of 1D-model and ASCOT for all the cases that were simulated are collected in Fig.~\ref{fig:allLossrates}. The loss rates are shown instead of particle density evolution for easier comparison.

\section*{References}
\bibliographystyle{iopart-num}
\bibliography{bibfile}

\end{document}